\documentclass[%
 aip,
 amsmath,amssymb,
 reprint,%
]{revtex4-1}

\usepackage{graphicx}
\usepackage{dcolumn}
\usepackage{bm}
\usepackage{placeins} 
\usepackage[utf8]{inputenc}
\usepackage[T1]{fontenc}

\usepackage{etoolbox}
\usepackage{booktabs}

\makeatletter
\def\@email#1#2{%
 \endgroup
 \patchcmd{\titleblock@produce}
  {\frontmatter@RRAPformat}
  {\frontmatter@RRAPformat{\produce@RRAP{*#1\href{mailto:#2}{#2}}}\frontmatter@RRAPformat}
  {}{}
}%
\makeatother
\begin{document}

\preprint{AIP/123-QED}

\title[Coarse-grained modelling of DNA-RNA hybrids]{Coarse-grained modelling of DNA-RNA hybrids}

\author{Eryk J. Ratajczyk}
 \affiliation{ 
Clarendon Laboratory, Department of Physics, University of Oxford, Parks Road, Oxford OX1 3PU, United Kingdom
}%
 \affiliation{ 
Kavli Institute for Nanoscience Discovery, University of Oxford,
Dorothy Crowfoot Hodgkin Building, South Parks Road, Oxford OX1 3QU, United Kingdom
}%
\author{Petr Šulc}%

\affiliation{ 
School of Molecular Sciences and Center for Molecular Design and Biomimetics,
The Biodesign Institute, Arizona State University,
1001 South McAllister Avenue, Tempe, AZ 85281, USA
}%

\affiliation{
School of Natural Sciences, Department of Bioscience, Technical University Munich, 85748 Garching, Germany.
}%

\author{Andrew J. Turberfield}
 \affiliation{ 
Clarendon Laboratory, Department of Physics, University of Oxford, Parks Road, Oxford OX1 3PU, United Kingdom
}%
 \affiliation{ 
Kavli Institute for Nanoscience Discovery, University of Oxford,
Dorothy Crowfoot Hodgkin Building, South Parks Road, Oxford OX1 3QU, United Kingdom
}%

\author{Jonathan P.K. Doye}

\affiliation{%
Physical and Theoretical Chemistry Laboratory, Department of Chemistry, University of Oxford, South Parks Road, Oxford, OX1 3QZ, United Kingdom
}%

\author{Ard A. Louis}

\affiliation{%
Rudolf Peierls Centre for Theoretical Physics, University of Oxford, 1 Keble Road, Oxford, OX1 3NP,
United Kingdom
}%

\date{\today}

\begin{abstract}
We introduce oxNA, a new model for the simulation of DNA-RNA hybrids which is based on two previously developed coarse-grained models---oxDNA and oxRNA. The model naturally reproduces the physical properties of hybrid duplexes including their structure, persistence length and force-extension characteristics. By parameterising the DNA-RNA hydrogen bonding interaction we fit the model's thermodynamic properties to experimental data using both average-sequence and sequence-dependent parameters. To demonstrate the model's applicability we provide three examples of its use---calculating the free energy profiles of hybrid strand displacement reactions, studying the resolution of a short R-loop and simulating RNA-scaffolded wireframe origami.
\end{abstract}

\maketitle

\section{\label{sec:level1}Introduction} 

\noindent  DNA (deoxyribonucleic acid)
and RNA (ribonucleic acid) are sufficiently similar that they can form  stable DNA-RNA hybrids \cite{Milman1967}.  In a biological context an important example of such hybrids is the R-loop, which forms when one of the strands in double-helical DNA is displaced by complementary RNA to create a hybrid duplex and an unpaired DNA strand \cite{Petermann2022}. \emph{In vivo}, short R-loops form during nuclear DNA replication by RNA primers, as well as during transcription when nascent RNA anneals to the DNA template inside an RNA polymerase\cite{Aguilera2017dna}. The formation of an R-loop is also necessary for the proper functioning of RNA-guided endonucleases in CRISPR-Cas systems where the guide RNA must fully hybridise with its DNA target for cleavage to take place \cite{Zhang2021,Pacesa_2022,Jiang2017}. Much longer R-loops (of the order of 1 kilobase) are formed during the replication of mitochondrial DNA and immunoglobulin class-switch recombination \cite{Aguilera2017dna}. R-loops play an important role in gene regulation. Errors in their formation and resolution can cause DNA damage, transcription elongation defects, hyper-recombination and genome instability~\cite{niehrs2020regulatory}, and they are also implicated in disease \cite{Rinaldi2021,Brambati2020}. Finally, DNA-RNA hybridisation underlies the action of antisense oligonucleotide (ASO) drugs, a therapeutic modality which has shown great promise in, for example, the treatment of neurological disorders\cite{DiFusco2019,Lee2013,Rinaldi2017}.

The specificity and predictability of Watson-Crick base-pairing also makes DNA and RNA excellent candidate materials for the design of synthetic self-assembled nanostructures, underpinning the growing field of nucleic acid nanotechnology \cite{Krishnan2019}. By simply annealing sets of strands with designed patterns of sequence complementarity, DNA has been used to assemble complex shapes \cite{Rothemund_2006,Douglas_2009}, dynamic nanomachines \cite{Yurke2000,Turberfield2003,Andersen_2009,Pumm_2022} and constructs with potential therapeutic and diagnostic applications \cite{Li2018,Sigl_2021,Benenson2004,Douglas2012}. Due to the presence of non-canonical interactions in RNA, its self-assembly is less well characterised. However, the field of RNA nanotechnology is also  advancing rapidly, with many examples of functional nanostructures and methods for their assembly \cite{Chworos2004,Geary2014,McRae_2023}. The design of nanostructures comprising DNA hybridised to RNA is under-explored, although interest is  increasing with exciting potential uses such as the delivery of therapeutic mRNA and artificial ribozyme fabrication \cite{Zhou_2021, Wu_2021, Parsons_2023}. 

Many different approaches have been developed to tackle the problem of nucleic acid modelling and simulation. Analytical mathematical models such as the worm-like chain (WLC) \cite{Marantan_2018}, which treats DNA or RNA as a semi-flexible polymer, can be useful if one is not concerned with details of the structure of the system. Classical molecular dynamics simulations that consider effective interactions between every atom have yielded useful insights into nucleic acid structure and dynamics, although they can only access microsecond timescales \cite{Galindo_Murillo_2019,_poner_2014,_poner_2018}. Quantum-chemical calculation is the most fine-grained computational technique used to study nucleic acids\cite{_poner_2013}, but this is usually limited to very small systems such as dinucleotides \cite{Ml_dek_2013}. Coarse-grained models, in which groups of atoms are represented as single particles, are a viable intermediate which offers a compromise between speed and detail \cite{Hafner_2019,Kmiecik_2016}. While many coarse-grained models of DNA and RNA have been developed \cite{Sun_2021,Denesyuk_2013, Reshetnikov_2017,Li_2021,Dawson_2016,Maffeo_2014}, modelling of hybrid systems, coarse-grained or otherwise, is relatively sparse, and is mostly limited to atomistic simulations \cite{Cheatham_1997,Noy_2005,Liu_2019}. Other examples include a mesoscopic model parameterised to reproduce melting temperatures \cite{de_Oliveira_Martins_2019} and an abstract model for R-loop formation \cite{Jonoska_2020}.

Here, we combine the most up-to-date versions of the models for DNA and RNA developed within the oxDNA framework\cite{Poppleton2023} to enable the simulation of DNA-RNA hybrids. The original average-sequence DNA model \cite{Ouldridge_2011} has  been extended to introduce sequence-dependent thermodynamic properties  \cite{_ulc_2012}, improved structural properties and salt-dependence \cite{Snodin_2015}. The same coarse-graining methodology has been used to develop an RNA model \cite{_ulc_2014,Poppleton2023}. A version of the DNA model with sequence-dependent structural and elastic properties is currently under development. The oxDNA family of models has seen tremendous success as tools for the study of nucleic acids and have improved our understanding of DNA and RNA origami\cite{Snodin_2016,Huang_2019,Benson_2018,Engel_2018,Torelli_2018, Engel_2018, snodin2019coarse,Torelli_2020} and strand displacement reactions\cite{_ulc_2015,Srinivas_2013}, as well as fundamental nucleic acid biophysics\cite{romano2013coarse,ouldridge2013dna,mosayebi2015force,matek2015plectoneme,schreck2015dna,Kriegel_2018,Nomidis_2019,Suma_2023,Lim_2022}. The introduction of our hybrid model to include DNA-RNA interactions will further expand the range of systems that can be simulated.

\section{\label{sec:level1}The model} 
\noindent Here we provide a brief overview of the previously developed models for DNA and RNA as well as  the introduction of new inter-strand interactions that enable the simulation of hybrids. We then describe in detail how the model was parameterised.
\subsection{\label{sec:level2}oxDNA and oxRNA}

\noindent In both oxDNA and oxRNA, nucleotides are treated as rigid bodies with interaction sites at the backbone and base. The models take a top-down coarse-graining approach---instead of attempting to exactly replicate the complex intermolecular forces between nucleotides, we use a series of simplified, physically plausible pairwise interactions which we then parameterise so that our model reproduces desired properties of the system. The oxDNA/oxRNA interaction potential takes the following form:

\begin{equation}
    \begin{gathered}
      U = \sum_{bonded} V_{bb} + V_{stck} + V_{exc}+\\    \sum_{nonbonded} V_{HB} + V_{crstck} + V_{cxstck} + V_{DH} + V'_{exc}.
\end{gathered}
\label{eqn:oxdna_potential_original}
\end{equation}

\begin{figure}
\includegraphics[scale=0.12]{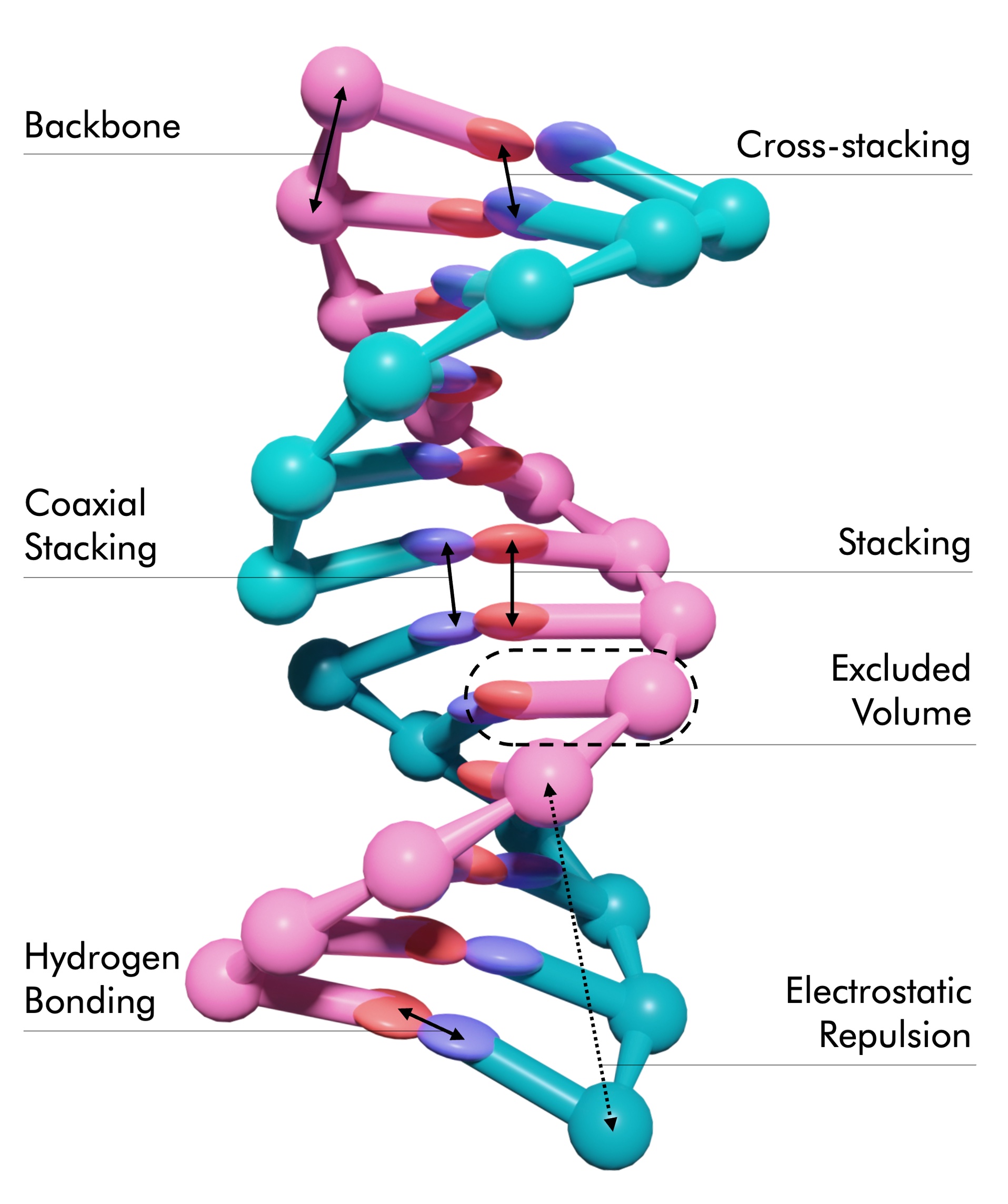}
\caption{\label{fig:model} A nucleic acid duplex as represented by the coarse-grained model, depicting DNA (blue) hybridised to RNA (pink). Interactions between nucleotides are indicated. Linkages between backbone sites indicate strand directionality, getting thinner in the $5'$ to $3'$ direction.}
\end{figure}

\noindent The first summation runs over all pairs of particles which are connected through covalent bonds, and includes $V_{bb}$ which enforces backbone connectivity, a stacking potential $V_{stck}$ and an excluded volume potential $V_{exc}$. The second summation runs over all remaining pairs of particles---which are not covalently bonded---and includes hydrogen bonding $V_{HB}$, cross-stacking $V_{crstck}$, coaxial stacking $V_{cxstck}$, a Debye-Hückel electrostatic interaction $V_{DH}$, and excluded volume $V'_{exc}$. Fig.\ \ref{fig:model} depicts how nucleic acids are represented in oxDNA and indicates the interactions between nucleotides. Coaxial stacking can be thought of as a modified version of the stacking interaction which is applied across a nick in a backbone. Both the bonded and non-bonded excluded volume terms are implemented as repulsive Lennard-Jones potentials between backbone-backbone, backbone-base and base-base interaction sites. Details of the exact functional forms of individual interactions can be found in the publications which first introduced the models \cite{Snodin_2015,_ulc_2014}. oxDNA and oxRNA operate within the same framework and differ only in the relative positions of interaction sites representing DNA/RNA nucleotides and the parameters which govern the strengths of interactions between them.

\subsection{\label{sec:level2}Incorporating DNA-RNA interactions}

\noindent The hybrid model was implemented by introducing a new DNA-RNA hybrid potential while using existing potentials to handle DNA-DNA and RNA-RNA interactions. The full interaction potential of our hybrid model, for a system containing both DNA and RNA, now reads

\begin{equation}
      U = U_{DNA} + U_{RNA} + U_{hybrid}.
      \label{eqn:hybrid_potential}
\end{equation}

\noindent
$U_{DNA}$ and $U_{RNA}$ include DNA-only and RNA-only interactions, respectively, and have the same general form as Eq. \ref{eqn:oxdna_potential_original}. Interactions between DNA and RNA are represented by $U_{hybrid}$, which uses the non-bonded inter-strand potentials---hydrogen bonding, cross stacking, coaxial stacking, the Debye-Hückel interaction and excluded volume. Since we do not allow covalent bonds between DNA and RNA nucleotides, $U_{hybrid}$ does not include any of the bonded interactions in Eq. \ref{eqn:hybrid_potential}. The forms of these new hybrid interactions are the same as those used in the DNA model and, unless otherwise specified, the same parameters were also used. 

\subsection{\label{sec:level2}Parameterisation}
\noindent Parameterisation of our hybrid model was constrained by the requirement to avoid changes to $U_{DNA}$ and $U_{RNA}$ in order to maintain compatibility with the original oxDNA and oxRNA models; only the hybrid DNA-RNA interactions were modified. In future versions it would be possible to reparameterise all of $U_{hybrid}$ and potentially obtain an even better fit to experimentally measured properties.

\begin{figure}
\includegraphics[scale=0.069]{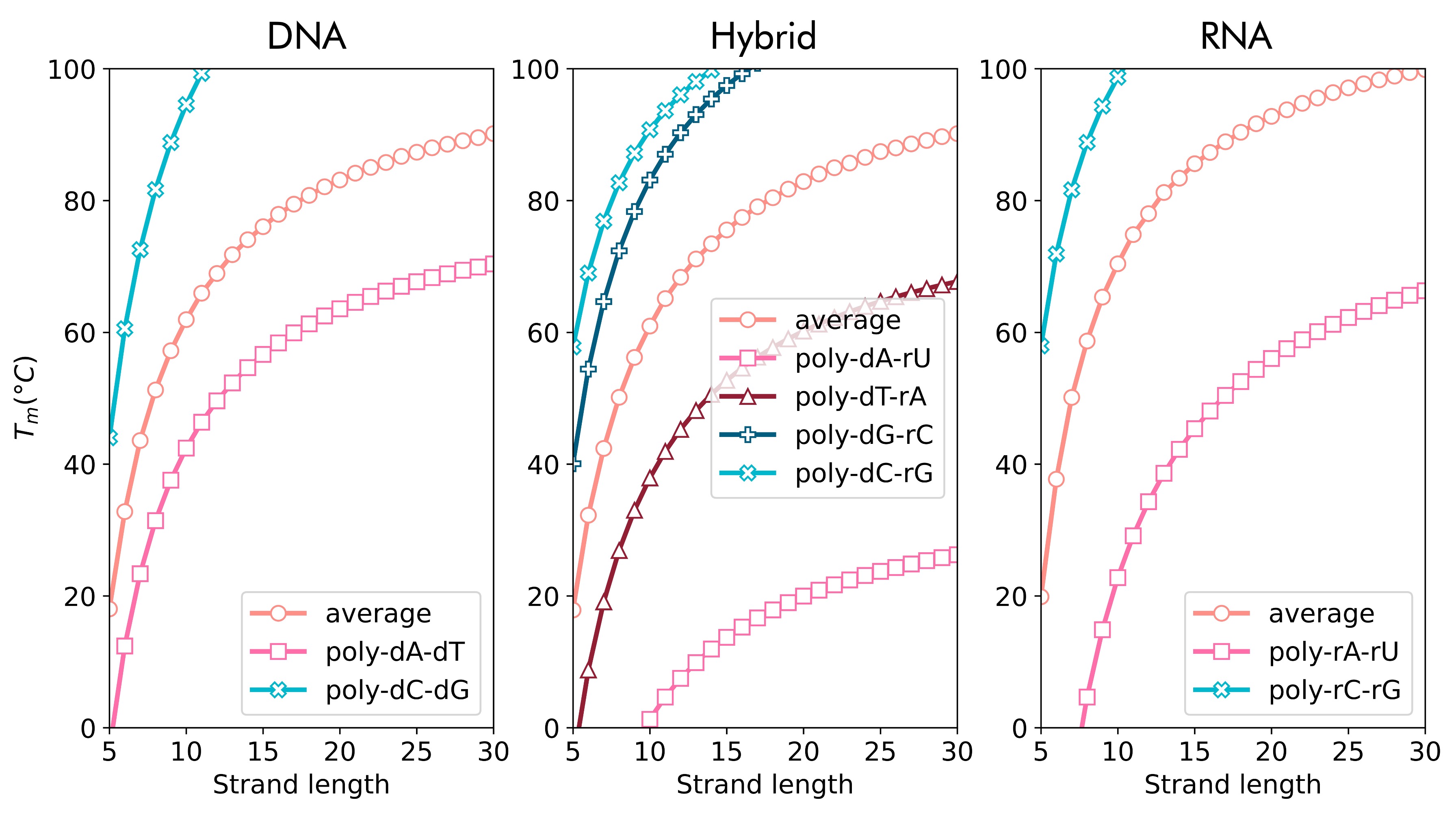}
\caption{\label{fig:sugimoto} Duplex melting temperature as a function of strand length for different sequences in DNA, DNA-RNA hybrids and RNA, as predicted by nearest neighbour models. For the average case, the mean melting temperature of 10\,000 random sequences was calculated.}
\end{figure}

As for oxDNA and oxRNA, we parameterise the hybrid model by fitting to the predictions of a nearest-neighbour model of thermodynamic properties which has itself been calibrated to reproduce experimental observations. Nearest-neighbour models for nucleic duplex formation are built by first conducting melting experiments for a range of sequences and using these data to estimate the thermodynamic parameters ($\Delta H$ and $\Delta S$) associated with the formation of every possible nucleotide pair in the context of its nearest neighbours (as well as initiation parameters). Using these parameters one can estimate the melting temperature ($T_m$) of an entire duplex, which is defined as the temperature at which the single-stranded (ss) and double-stranded (ds) states are equally probable. 

Sugimoto \textit{et al}.\ first estimated nearest-neighbour thermodynamic parameters for DNA-RNA hybrids over two decades ago\cite{Sugimoto_1995}. A more recent set of improved parameters (now also with sequence-specific initiation parameters) \cite{Banerjee_2020} is employed here. The Sugimoto nearest-neighbour model (SNN) predicts melting temperatures to an accuracy of roughly 1\,°C: for the purposes of this work we consider it to be a very good fit to experiment.  Fig.\ \ref{fig:sugimoto} highlights the drastic effect that sequence can have on melting temperature in DNA-RNA hybrids, also showing the differences in melting thermodynamics between hybrids, dsDNA and dsRNA. We used the nearest neighbour model of SantaLucia and Hicks \cite{SantaLucia2004} to estimate dsDNA melting temperatures, and the model of Xia \textit{et al}.\ for dsRNA\cite{Xia1998}. In both cases we employed an empirical salt correction to $T_m$ derived by SantaLucia\cite{SantaLucia1998}. Melting temperatures were calculated using Biopython 1.75\cite{Cock2009}. For hybrids it is also noteworthy that, for case of G-C base pairs, there is a dependence on not only sequence but also on the distribution of bases between the DNA and RNA strands, as illustrated by the difference between melting curves for poly-dG-rC and poly-dC-rG. The $T_m$ shown are at a monovalent salt concentration of 1\,M, and a total strand concentration of $3.5\times 10 ^{-4}$\,M---values which were also used in melting simulations.

To parameterise our model we selected hydrogen bonding strength parameters (i.e.\ potential well depths) for hybrid A-U, A-T and G-C base-pairs that reproduce the melting temperatures predicted by the Sugimoto model. To estimate the melting temperatures of hybrid duplexes predicted by the model, we simulated duplex dynamics near the melting temperature using the Virtual Move Monte Carlo (VMMC) algorithm \cite{Whitelam_2007} and umbrella sampling \cite{Torrie_1977}.  Umbrella sampling weights were chosen to ensure that the transition between the single- and double-stranded state was thoroughly sampled. For any given duplex, simulations were run for $10^9$ time-steps (three independent simulations for average-sequence, one for sequence-dependent versions of the model) at the melting temperature predicted by the Sugimoto model. From these simulations we obtain the equilibrium populations of single- and double-stranded states and extrapolate to the temperature at which these states are equally probable---details of the method can be found in the paper introducing oxRNA \cite{_ulc_2014}. We sought a set of parameters which minimise the following cost function:
\begin{gather}
    C = \sum_{i \in S} \Delta T_m(i)^2,
\end{gather}
where $\Delta T_m(i) = T_m^{VMMC}(i) - T_m^{SNN}(i)$, $T_m^{VMMC}(i)$ and $T_m^{SNN}(i)$ are the melting temperatures predicted by VMMC simulations using our model and the SNN model, respectively, for a given sequence $i$. The sum runs over a training library, $S$. For the sequence-dependent parameterisation, $S = S_{dep}$, comprising data from 4096 6-mers and 30\,000 each of random 8-, 10- and 12-mers. In the average-sequence parameterisation strand length is the only determinant of $T_m$, in which case we use $S = S_{avg}$, which only contains four training points: strands of length 6, 8, 10 and 12.

\begin{figure}
\includegraphics[scale=0.115]{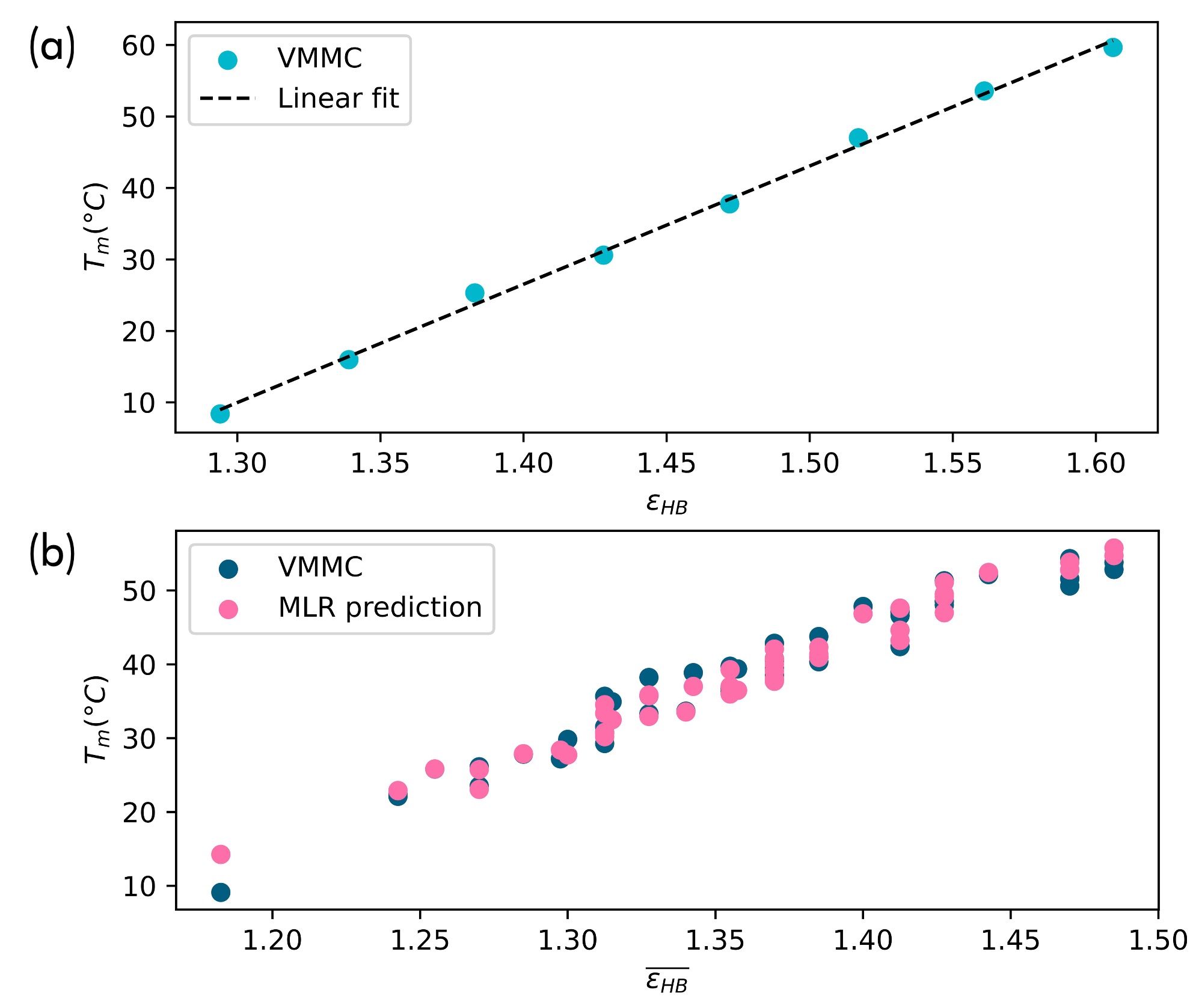}
\caption{\label{fig:seq_tm_mapping} Dependence of melting temperature of an 8-mer on hydrogen bonding strength, obtained from VMMC simulations. (a) Melting temperature as a function of hydrogen bonding strength calculated using the average-sequence model alongside a fitted straight line. (b) Melting temperatures of 50 random sequences, calculated by VMMC simulation, plotted against the mean hydrogen bonding strength of the sequence, and the corresponding fits to an MLR model.}
\end{figure}

For the average-sequence model we assume that all hydrogen bonds have the same strength, $\varepsilon_{HB}$. We ran melting simulations for a range of bond strengths for duplexes of length 6, 8, 10 and 12---in each case we find a linear relationship between $T_m^{VMMC}(i)$ and $\varepsilon_{HB}$. For each duplex length we fit a straight line to the data (Fig.\ \ref{fig:seq_tm_mapping}(a)), enabling accurate prediction of $T_m^{VMMC}(i)$ from $\varepsilon_{HB}$. We chose the value of $\varepsilon_{HB}$ which minimises $C$ over the average-sequence training library, $S_{avg}$.

In the sequence-dependent case we have four possible types of hydrogen bond, thus four parameters to select: $\varepsilon_{dArU}$, $\varepsilon_{dTrA}$, $\varepsilon_{dGrC}$ and $\varepsilon_{dCrG}$. In order to fit sequence-dependent parameters, previous iterations of our coarse-grained models used a histogram reweighting technique to calculate melting temperatures and an annealing algorithm to search the parameter space \cite{_ulc_2012,_ulc_2014,Snodin_2015}. This approach is necessary when one is fitting >10 parameters, many of which, e.g.\ stacking strengths, have quite subtle effects on $T_m$. Here, since the parameter space to be searched is much smaller, we are able to use a simpler method. We first find an approximate linear mapping between sequence and the melting temperature predicted by our model. We then use this mapping to find the parameters which best reproduce melting temperatures predicted by the Sugimoto model. 

In order to search the parameter space, we used the following initial minimisation procedure: (1) Initialise parameters to average-sequence values. (2) For every sequence in $S_{dep}$, use the current values of $\varepsilon_{dXrY}$ to calculate the average bond strength $\overline{\varepsilon}_{HB}(i)$ and, assuming the linear scaling established for the average-sequence model, the corresponding approximation to $T_m^{VMMC}(i)$. Compute $C$. (3) Randomly perturb parameters to generate new parameters, and repeat step 2 for the new parameter set. (4) If new parameters reduce $C$, accept them, otherwise repeat step 3. (5) Repeat steps 2 to 4 until $C$ converges. 

In order to further refine the parameters, we needed a better mapping between sequence and $T_m^{VMMC}(i)$. With this in mind, we used multiple linear regression (MLR) to predict the result of a VMMC calculation of the melting temperature of a sequence $i$ of length $k$, such that
\begin{gather}
    T_m^{VMMC}(i) = \beta_0 + \beta_1 x_1(i) + ... + \beta_k x_k(i),
\end{gather}
where $x_n(i)$ is the hydrogen bonding strength of the $n^{th}$ base-pair within the duplex (read out in a $5'$ to $3'$ direction with respect to the DNA strand), and $\beta_0, \beta_1,...,\beta_k$ are fitting parameters obtained from a least-squares minimisation. For example, for a hybrid duplex $5'-$dATGC$-3'$/$3'-$rUACG$-5'$, we would estimate $T_m$ as $\beta_0 + \beta_1 \varepsilon_{dArU} + \beta_2 \varepsilon_{dTrA} + \beta_3 \varepsilon_{dGrC} + \beta_4 \varepsilon_{dCrG}$. 

Using our previously obtained estimates of the bonding parameters, we ran melting simulations for 500 random sequences (125 per duplex length) and used these data to fit an MLR model for each length of duplex (Fig.\ \ref{fig:seq_tm_mapping}(b)). We then performed the minimisation procedure described above---now with the improved VMMC predictions made using the MLR models---to arrive at a refined parameter set. We repeated all of the above (i.e.\ melting simulations of 500 random sequences, MLR model fitting, followed by the minimisation) for a final time, but found that by this point the parameters had converged.

Note that initially we also attempted to fit the strength of the cross-stacking interaction for the average-sequence model. We performed preliminary fitting of the hydrogen bonding and cross-stacking ($K^{hybrid}_{crstck}$) strengths simultaneously, and found that the value which minimised $C$ was $K^{hybrid}_{crstck} = 0.938K^{DNA}_{crstck}$, where $K^{DNA}_{crstck}$ is the value used by the DNA model (for reference, $K^{RNA}_{crstck} = 1.262K^{DNA}_{crstck}$). However, we also found that increasing $K^{hybrid}_{crstck}$ while decreasing $\varepsilon_{HB}$ accordingly (and vice versa), made little difference to the overall fit. Since the relative strengths of the cross-stacking and hydrogen bonding interactions are not experimentally constrained, different values for $K^{hybrid}_{crstck}$ and $\varepsilon_{HB}$ could have been chosen without detriment to the model. We chose to set $K^{hybrid}_{crstck} = 0.938K^{DNA}_{crstck}$ and then selected $\varepsilon_{HB}$ using the procedure outlined in Section II C.

The coaxial stacking and Debye-Hückel interactions also could, in principle, have been reparameterised. However, to the best of our knowledge, no data for DNA-RNA hybrids exists which could be used to fit these interactions. We set parameters of Debye-Hückel interaction for hybrids to the values used by the RNA model. The hybrid model uses the same coaxial stacking interaction as the DNA model. 

\begin{figure*}
\includegraphics[scale=0.11]{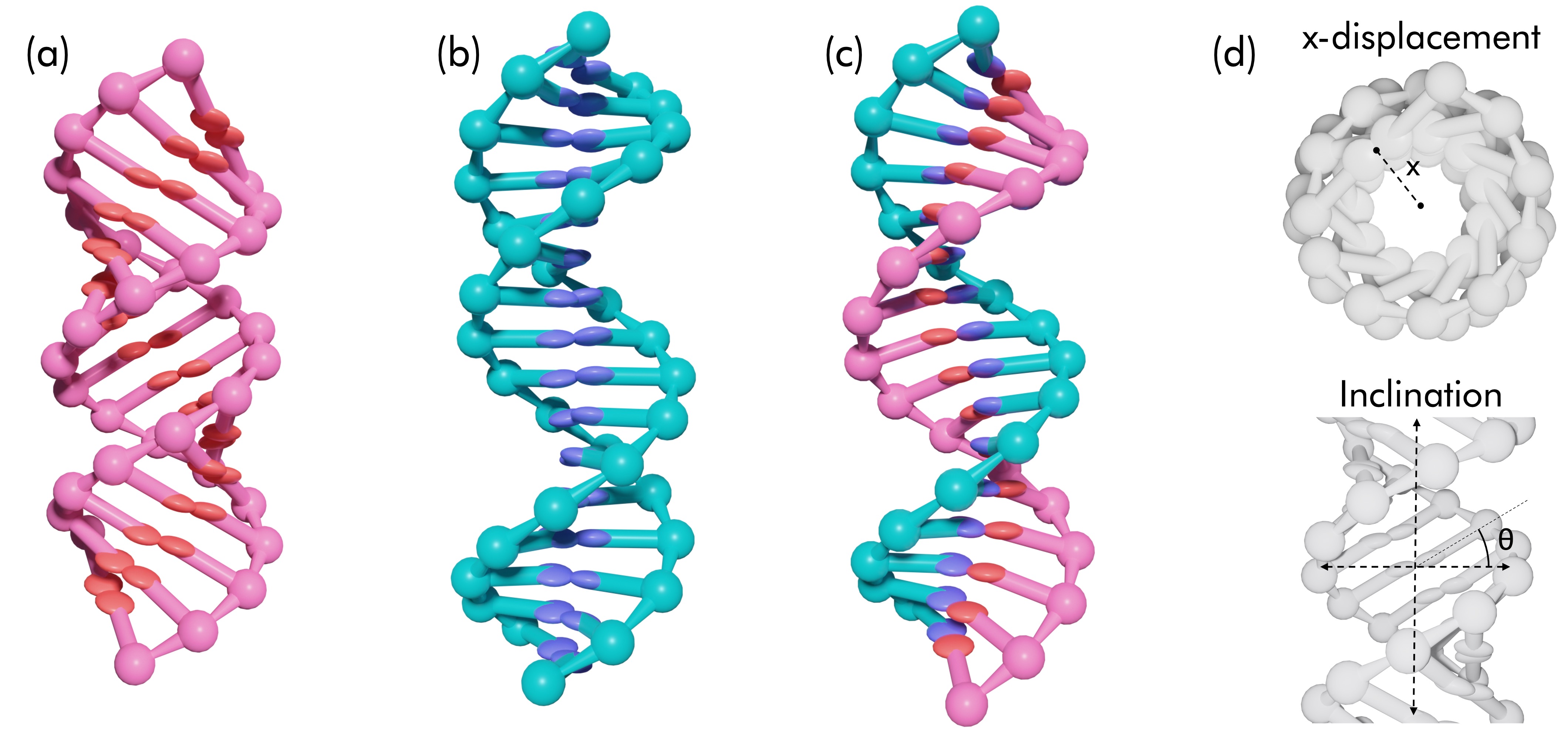}
\caption{\label{fig:structure} A comparison of the structures of double-stranded nucleic acids in oxDNA and oxRNA. Shown side-by-side are the structures adopted by a 16-mer of (a) RNA, (b) DNA and (c) a DNA-RNA hybrid, which naturally adopts a conformation somewhere between that of A-form RNA and B-form DNA. (d) An illustration of the x-displacement and inclination helical parameters, which are used to structurally characterise the helices. The shortest distance, $x$, between the helical axis and the interaction site where bases meet is defined as x-displacement. Inclination is the angle made between a base and the plane perpendicular to the helical axis, indicated as $\theta$.}
\end{figure*}

\section{\label{sec:level1}Properties of the Model}
\noindent In this section we report the physical predictions of our model. These include the structure of double-stranded DNA-RNA hybrid duplexes, the melting behaviour of both the average-sequence and sequence-dependent versions of our model, and  mechanical properties such as persistence length and force-extension characteristics.

\subsection{\label{sec:level2}Structure}
\noindent The structures of double-stranded DNA and RNA differ significantly---DNA most commonly folds into a B-form helix, whereas RNA takes up an A-form conformation. The A-form helix is characterized by significant slide (displacement of adjacent base pairs along the long axis of the pair) and roll (the angle by which base-pairs open up toward the minor groove), with the result that base pairs are shifted away from the helical axis and inclined to it\cite{Calladine1984,Dickerson_2001}. Fig.\ \ref{fig:structure}(d) defines the parameters x-displacement and inclination\cite{Hartmann1996} which are used to characterise the structure of the double helix in this work.

Reports on the exact structure of DNA-RNA hybrids vary. Thanks to studies of polymeric hybrids it is largely accepted that poly-rA-dT can experience an A- to B-form transition with changes in relative humidity \cite{Shaw2008}. Hybrids containing poly-dA-rU or poly-dI-rC have been termed \textit{heteromerous}, whereby the DNA and RNA strands possess B- and A-form characteristics respectively\cite{Arnott1986}. The detailed structure of oligomeric hybrids can depend on sequence---it is known, for instance, that the purine/pyrimidine content of the DNA strand can change the backbone conformation \cite{Wheelhouse_2009}. An NMR study by Gyi \textit{et al}.\ \cite{Gyi1998} found that the extent of A- or B-form helicity, as well as the major/minor groove widths vary with purine/pyrimidine content. They also found that a high-purine DNA strand results in greater conformational diversity as a result of increased sugar flexibility, compared to the case when the RNA strand of the hybrid duplex is high in purine. More recent crystallography studies of oligomeric DNA-RNA hybrids typically characterise them as A-form \cite{Conn_1999,Cofsky_2022,Xiong_2000}, and estimates of their exact x-displacement and inclination obtained from all-atom simulations suggest a structure in-between those of DNA and RNA \cite{Liu_2019}.

To determine the structure of a hybrid duplex in our model, and compare it to DNA and RNA, we generated 10\,000 uncorrelated configurations of each class of 16 base-pair duplex by performing average-sequence Monte Carlo simulations at 25\,°C, with a monovalent salt concentration of 0.5\,M. We then measured the helical parameters of each configuration and calculated their means, shown in Table I. Note that the values for the RNA model differ from those first reported by Šulc \textit{et al}.\ \cite{_ulc_2014}, since the model used here includes salt-dependent effects which were not included in the original model. Representative structures are shown in Fig.\ \ref{fig:structure}. We see that the values of inclination, x-displacement and pitch are intermediate with respect to those for DNA and RNA. While our coarse-grained models are not primarily designed to achieve structural accuracy, it is encouraging that the high-level structural features of DNA-RNA hybrids emerge without being explicitly imposed.

In oxRNA, an A-form conformation is imposed on the helix by making the stacking interaction dependent on the angle between the  nucleotide orientation vector and the backbone vector connecting neighbouring nucleotides, such that the potential energy of the stacking interaction is minimised if the helix adopts an A-form geometry. This angular dependence is not present in the DNA model and, in hybrids, only the RNA strand has this modified stacking interaction. However, the short range of the hydrogen-bonding interaction forces base pairs to lie approximately in the same plane, resulting in a compromise between A- and B-forms. We find that this intermediate helix geometry has an effect on thermodynamic propertiess which is discussed in the next section.

\begin{table}[]
\begin{ruledtabular}
\begin{tabular}{@{}lllll@{}}

\textbf{Parameter} & \textbf{DNA} & \textbf{Hybrid} & \textbf{RNA} &  \\ \midrule
Inclination (°)      & 5.15   & 8.31   & 13.8   &  \\
x-displacement (nm)    & 0.0536   & 0.265   & 0.549   &  \\
Pitch (bp/turn)    & 10.6 \footnote{As reported by Snodin \textit{et al}.\ \cite{Snodin_2015}}   & 10.8   & 11.0   &  \\ 
Rise (nm/bp)    & 0.347 \footnote{As reported by Snodin \textit{et al}.\ at 0.5\,M salt\cite{Snodin_2015}}   & 0.343   & 0.280 \footnote{As reported by Šulc \textit{et al}.\ for the first version of oxRNA \cite{_ulc_2014}} &  \\
\end{tabular}
\caption{Comparison of the inclination, x-displacement, pitch and rise helical parameters for double stranded nucleic acids, obtained from simulations of our model.}
\end{ruledtabular}
\end{table}

\subsection{\label{sec:level2}Thermodynamics}

\begin{table}[]
\begin{ruledtabular}
\begin{tabular}{@{}lllll@{}}

\textbf{Parameter} & \textbf{DNA} & \textbf{Hybrid} & \textbf{RNA} &  \\ \midrule
$\varepsilon_{HB}$      & 1.07   & 1.50   & 0.87   &  \\
$\varepsilon_{dArU}$    & N/a   & 1.21   & 0.82   &  \\
$\varepsilon_{dTrA}$    & 0.89  & 1.37   & N/a   &  \\ 
$\varepsilon_{dGrC/dCrG}$   & 1.23  & 1.61/1.77   & 1.06   &  \\ 
\end{tabular}
\caption{The hydrogen bonding parameters of the model (in simulation units), compared to analogous parameters for the DNA and RNA models.}
\end{ruledtabular}
\end{table}

\noindent The model parameters selected by the fitting procedure described in Section II C, and used below, are shown in Table II. We note that the hydrogen bonding parameters required to reproduce the correct melting temperatures are substantially larger than in either oxDNA or oxRNA: this point is discussed below.  

\begin{figure}
\includegraphics[scale=0.28]{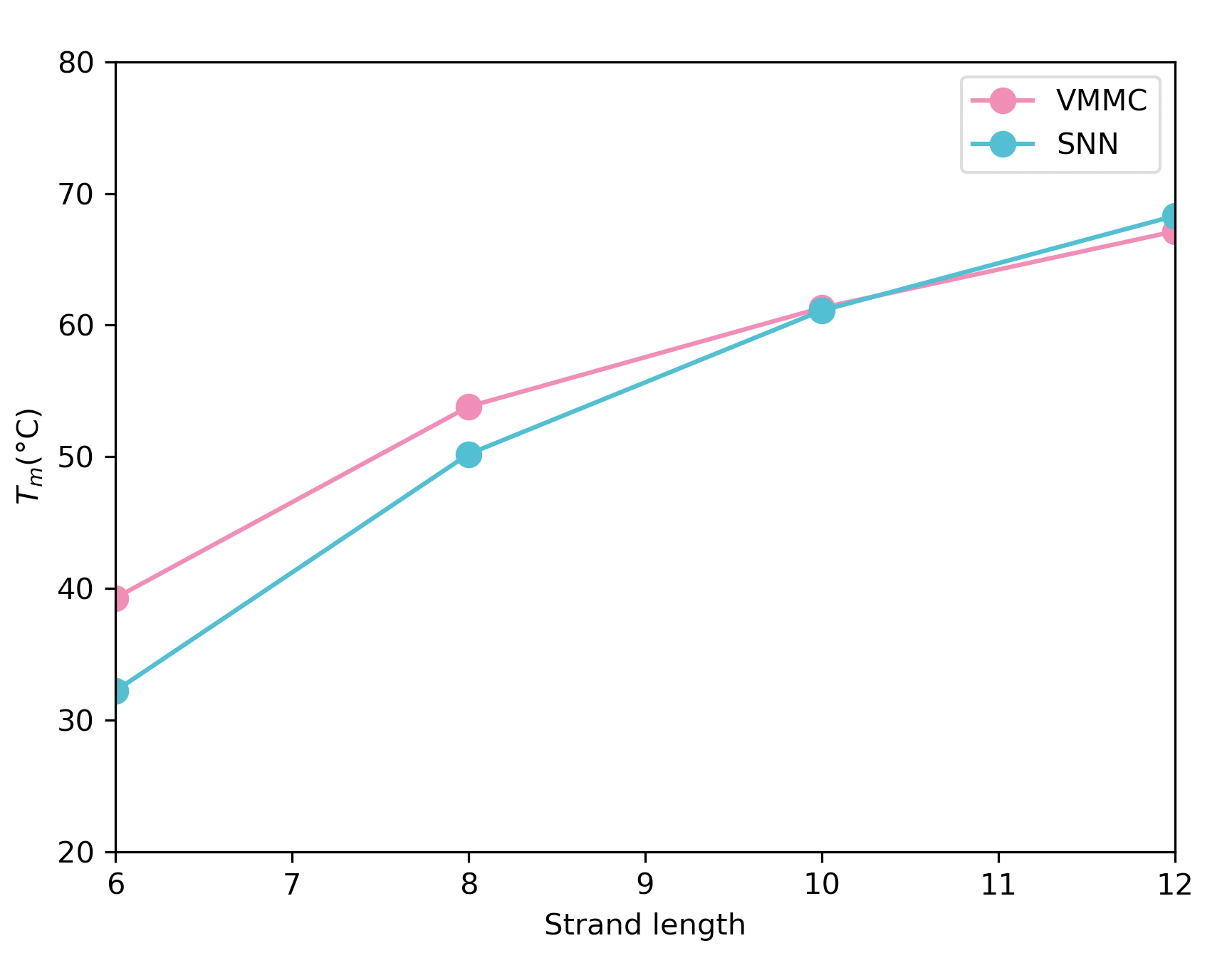}
\caption{\label{fig:average_melt} Melting temperature as a function of duplex length calculated for the average-sequence hybrid model, using VMMC simulations, compared to the target $T_m$ obtained from the Sugimoto nearest neighbour model.}
\end{figure}

The fit of the average-sequence model to target melting temperatures is shown in Fig.\ \ref{fig:average_melt}. While in general our model reproduces the melting behaviour of short hybrid duplexes quite well, there is a noticeable deviation from target temperatures at short strand lengths---for strands of length 6 and 8 the melting temperature is overestimated by around 7.1\,°C and 3.6\,°C, respectively. In the average-sequence DNA and RNA models corresponding deviations are typically no more than 1\,°C.

In order to investigate how hybridisation between DNA and RNA affects individual interactions, we computed the mean potential energies associated with stacking and hydrogen bonding using a simulation protocol similar to that used in Section III A but with the temperature set to 1\,°C in order to reduce fluctuations away from the double-stranded ground state. In general, stacking contributes less to the stability of hybrids than of dsDNA or dsRNA duplexes. This is because A- and B-form geometries respectively were imposed onto the RNA and DNA models through the forms of the interaction potentials: when part of a hybrid duplex, neither the DNA nor RNA is in its preferred conformation, which has a destabilising effect. This explains why the fitting procedure described in Section II C increases the hydrogen bonding strengths to compensate (cf. Table II). We also find that, as strand length increases, both stacking and hydrogen bonding interactions become, on average, less stabilising. This can be understood as a consequence of stabilising relaxation of the strained duplex near the ends---which becomes relatively less important as the duplex increases in length. It is also noteworthy that stacking is more disrupted for the RNA strand of a hybrid duplex than for the DNA strand. We propose that this tendency for the (RNA) stacking and hydrogen bonding to weaken with increasing strand length is the reason for the melting temperature overestimation in 6- and 8-mers. The model could be further adapted to include a modified stacking potential which can better accommodate hybrids, enabling an even better fit to experimental melting temperatures. This could be implemented by including a double-well angular/radial dependence in the stacking interactions, such that A- and B-form helicities
are maintained in dsRNA and dsDNA respectively, while
also allowing a hybrid duplex to inhabit a second potential energy well, mitigating the destabilising effect in the current version of the model.

\begin{figure}
\includegraphics[scale=0.7]{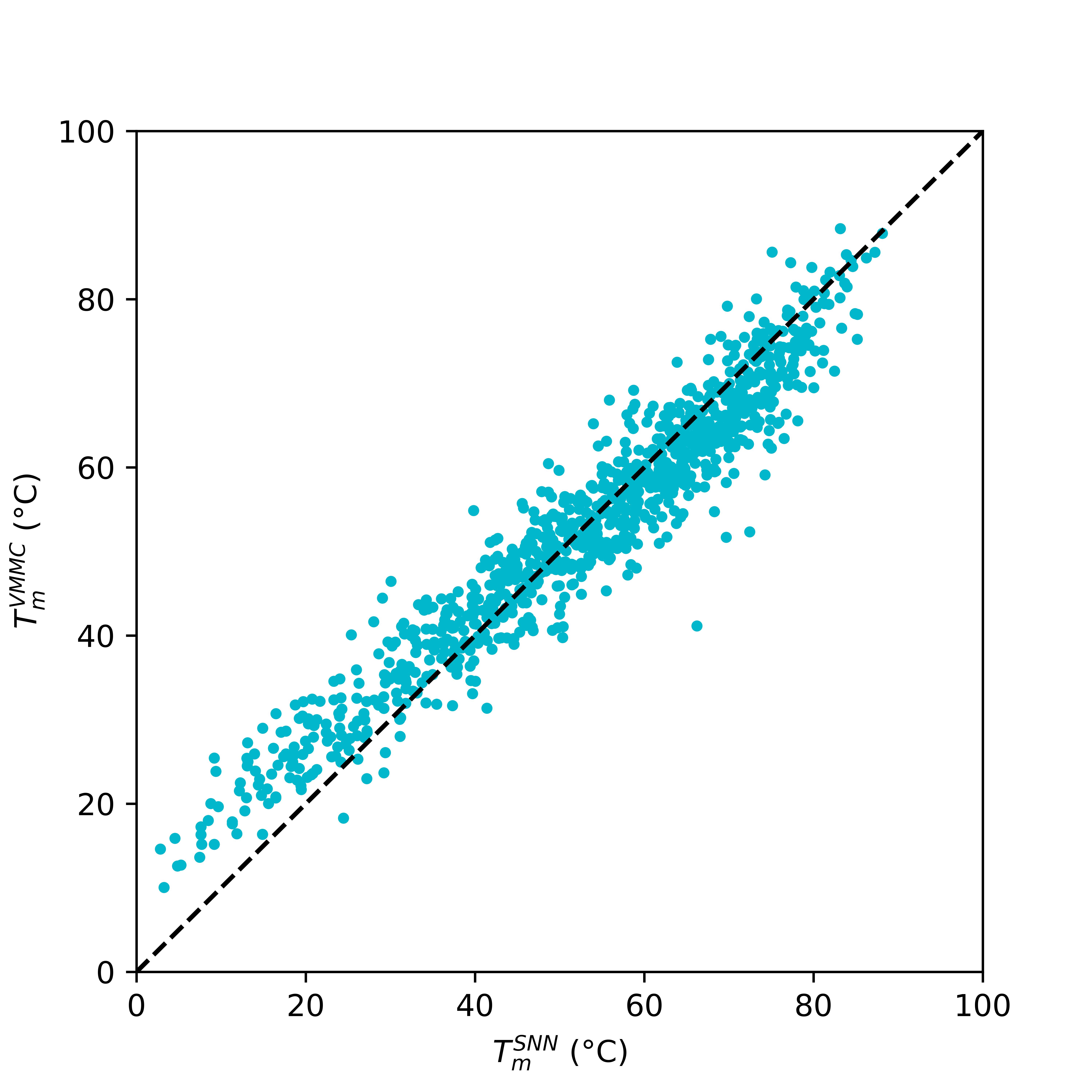}
\caption{\label{fig:seq_dep_melt} Performance of the sequence-dependent hybrid model, tested on 1000 random sequences. The plot shows the melting temperature predicted by the model, against the value predicted by the Sugimoto nearest neighbour model. The dashed line indicates $y=x$.}
\end{figure}

In order to test the sequence-dependent version of the model, we ran melting simulations on 1000 random duplexes of lengths 6, 8, 10 and 12 (250 per length). Sequences with predicted melting temperatures below 1\,°C (short, U-rich sequences) were discarded. Results are shown in Fig.\ \ref{fig:seq_dep_melt}. 
Over this 1000-sequence test set, the model achieves a mean $\Delta T_m$ of 0.0926\,°C, with a standard deviation of 5.36\,°C. While we consider this to be a more than satisfactory fit, we are aware of factors which limit our model's performance. The first is its over-estimation of the stability of short duplexes, as discussed for the average-sequence model. In Fig.\ \ref{fig:seq_dep_melt}, there is noticeable overestimation of $T_m$ in the <30\,°C region, which is almost certainly a manifestation of this effect. As discussed in the previous section, sequence can affect backbone conformation. Our model does not factor in these structural changes, which likely worsens the overall sequence-dependent fit.

\subsection{\label{sec:level2}Mechanical properties}

\noindent The mechanical properties of nucleic acids are biologically important \cite{Marin_Gonzalez_2017} and determine the mechanical behaviour of synthetic constructs like DNA origami \cite{Ji_2021}. For this reason, it is important to check that our model captures the basic mechanics of double-stranded DNA-RNA hybrids. Here we measure the persistence length and force-extension characteristics of hybrid duplexes within our model, and compare the results to available experimental data.

\begin{figure}
\includegraphics[scale=0.12]{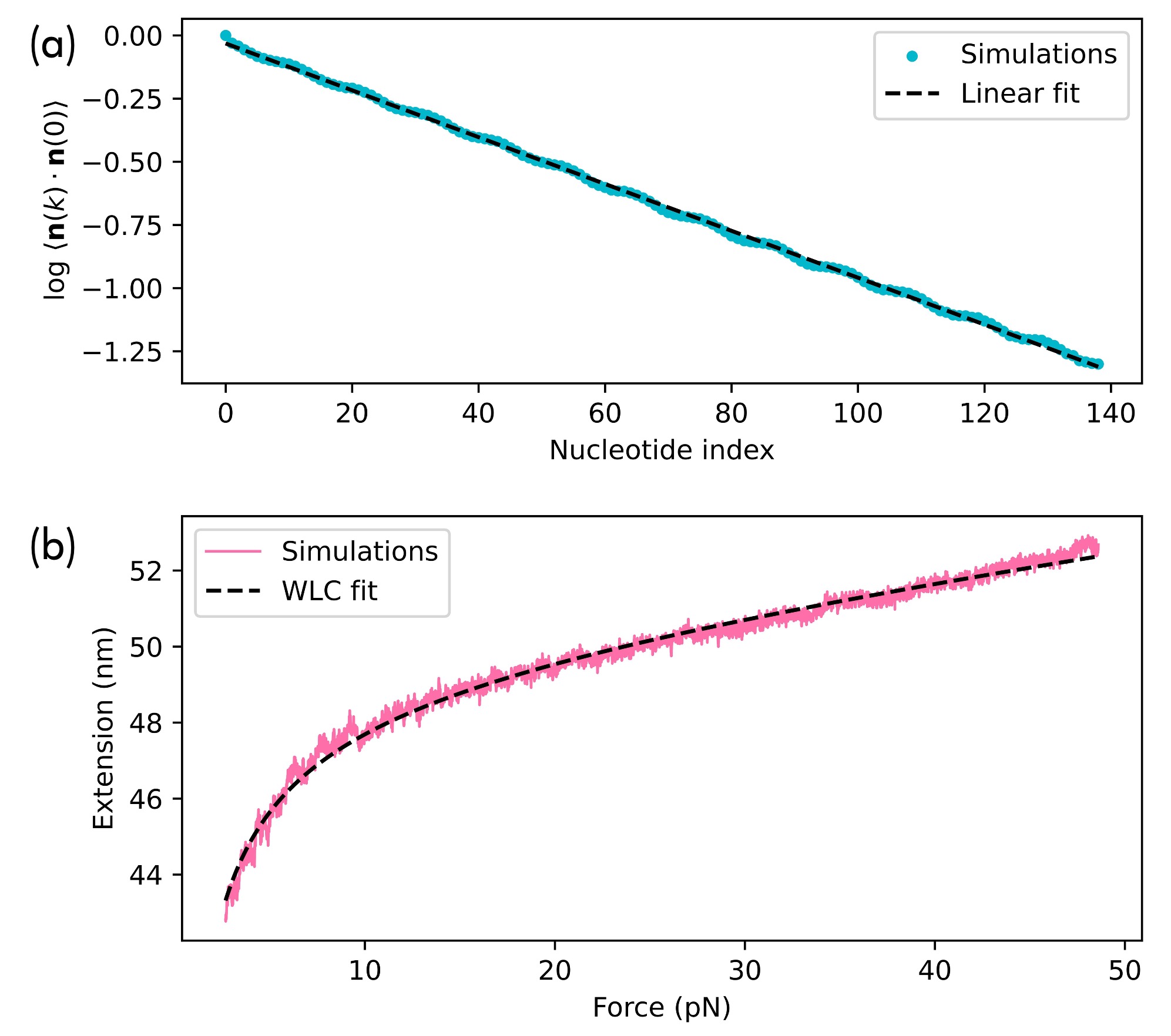}
\caption{\label{fig:mechanics} Measuring the mechanical properties of a 150-mer DNA-RNA hybrid. (a) Natural logarithm of the correlation function $\langle \textbf{n}(k) \cdot \textbf{n}(0) \rangle $ against the nucleotide index, $k$. (b) The force-extension curve obtained from simulations, alongside a fit to the extensible worm-like chain.}
\end{figure}

The persistence length $L_p$ of a polymer quantifies its bending stiffness. In a semi-flexible, infinitely long polymer, the persistence length quantifies the correlation between local helix orientations: 
\begin{equation}
    \langle \textbf{n}(k) \cdot \textbf{n}(0) \rangle = \exp \left (\frac{-k\langle r \rangle}{L_p} \right),
\end{equation}
where $\textbf{n}(k)$ is the local helical axis vector of the $k^{th}$ base-pair along the duplex and $\langle r \rangle$ is the rise per base-pair\cite{Doi1988-ky}. To measure $L_p$, we performed molecular dynamics (MD) simulations of a 150 base-pair hybrid duplex with the average-sequence model at 22\,°C, with the monovalent salt concentration set to 0.5\,M. We ran 10 independent simulations, each for $10^8$ time-steps, integrated using Langevin dynamics with a damping constant equal to the time-step. We sampled simulation frames every $10^4$ time-steps giving us a total of $10^5$ configurations. For each base-pair we computed the centre of mass, and translated it to account for the shift in an A-form helix to give us a point on the helical axis. From these points we calculate local helical axis vectors which are used to obtain $ \langle \textbf{n}(k) \cdot \textbf{n}(0) \rangle$. We discard the 5 terminal base-pairs to avoid end effects. From the gradient of the line in Fig.\ \ref{fig:mechanics}(a), we obtain an estimate of $L_p$ = 39\,nm.

A separate set of simulations was performed to measure the force-extension relationship. We used the same settings as before, except that in this case we ran 30 replicas, each for $10^7$ time-steps. We applied a uniformly increasing, equal and opposite force of up to 50\,pN to terminal nucleotides and sample the distance between them every $10^3$ time-steps to measure the extension, which was averaged over independent simulations. In this case, we fit our data to the extensible worm-like chain model\cite{Odijk_1995} which predicts that the projected end-to-end distance $L$ of a polymer along the direction of a force with magnitude $F$ is, in the limit $F > k_B T / 2 L_p$,

\begin{equation}
    L = L_c \left[  1 + \frac{F}{K} - \frac{k_B T}{2 F L_c}(1 + A\coth A)\right],
\end{equation}
where 
\begin{equation*}
    A = \sqrt{\frac{FL_c^2}{L_p k_B T}},
\end{equation*}
$K$ is the stretching modulus, and $L_c$ is the relaxed contour length. The results are shown in Fig.\ \ref{fig:mechanics}(b). Fitting to our data gives $K$ = 780\,pN, $L_c$ = 51\,nm and $L_p$ = 20\,nm. It must be pointed out that the value of $K$ especially is quite sensitive to the size of the fitting window---for example, fitting up to only 30\,pN doubles the estimated stretching modulus ($L_p$ is 25\% lower and $L_c$ changes very little). Note also that a similar issue was observed for oxRNA (but not oxDNA) and was ascribed to a decrease in the inclination angle as the force increased\cite{_ulc_2014}. Consequently, the error on these estimates can be assumed to be relatively large, which should be kept in mind when comparing to experimental values, and the persistence length obtained from the tangent-tangent correlation function should be considered to be more accurate. 

Experimental data on the mechanics of hybrid duplexes is scarce, and the number of all-atom simulation studies is also low. Zhang \textit{et al}.\ \cite{Zhang_2019} performed a series of magnetic tweezer experiments to measure the mechanical properties of a long (>10 kilobase) hybrid duplex at different salt concentrations. They report a stretching modulus of 660\,pN, which does not depend strongly on salt concentration. Conversely, salt does have an effect on persistence length which ranges from 49\,nm to 63\,nm at salt concentrations of 0.5\,M and 1\,M respectively. An all-atom simulation study performed at 1\,M monovalent salt concentration estimated a stretching modulus of 834\,pN \cite{Liu_2019}. Given that the model is parameterised to reproduce thermodynamic properties, the agreement between calculated and measured elastic properties is satisfactory. We note that for low applied forces (<35\,pN or so) the persistence length is more significant than the stretching modulus in determining the mechanical behaviour of the duplex. 

To put this into perspective, the persistence length $L_p$ of dsDNA at moderate to high salt concentration is in the range 45--50\,nm, and the stretching modulus $K$ is around 1050--1250\,pN at high salt \cite{Ouldridge_2011}. The first version of the oxDNA model achieves $L_p$ = 43.8\,nm and $K$ = 2120\,pN. For dsRNA, experimental estimates of $L_p$ are in the range 58--80\,nm and $K$ = 615\,pN, while for the oxRNA model $L_p$ = 28.3\,nm and $K$ = 296\,pN.




\section{\label{sec:level1}Applications of the model} 

\noindent We provide examples of the application of the model to three hybrid systems---toehold-mediated strand displacement, a short R-loop and RNA-scaffolded wireframe origami, all of which are technologically and/or biologically important.

\subsection{\label{sec:level2}Toehold-mediated strand displacement}

\begin{figure*}
\includegraphics[scale=0.12]{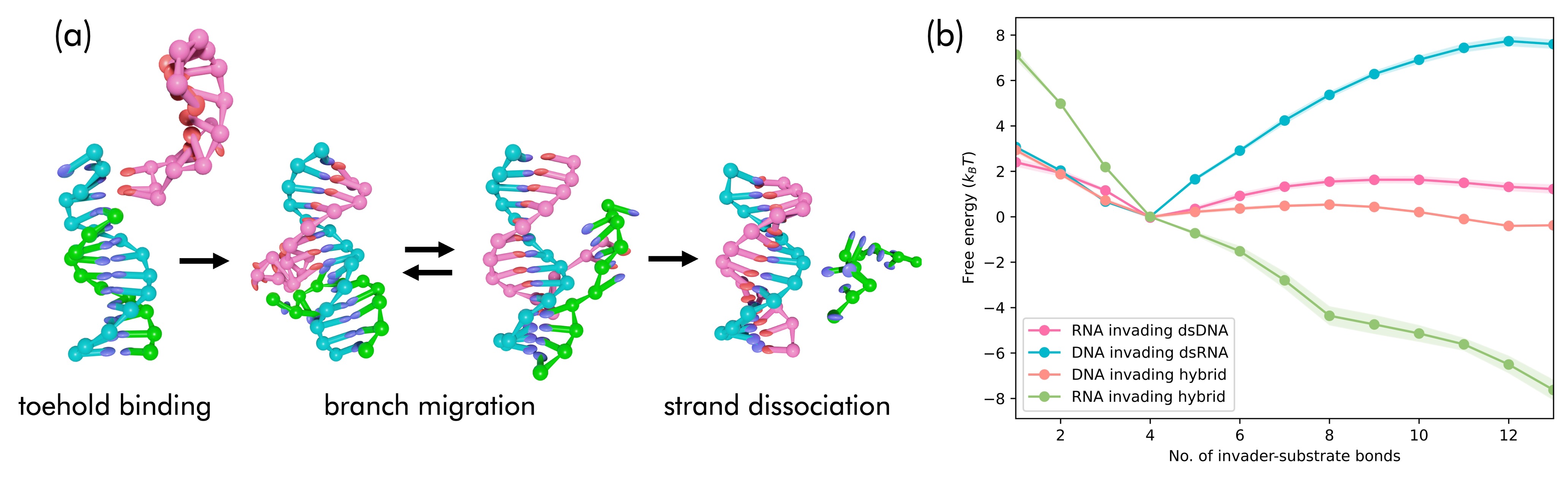}
\caption{\label{fig:tmsd} Simulation of toehold-mediated strand displacement using the average-sequence model. (a) Snapshots of key steps for one of the strand displacement reactions: An RNA strand (pink) invades a DNA duplex (blue and green) by binding initially to a single-stranded toehold. (b) Free energy profiles of the different systems simulated, showing the free energy (set to zero for a fully-occupied toehold) against the number of hydrogen bonds between the substrate and invader strands. Error bars indicate the standard error of the mean.}
\end{figure*}

\noindent Toehold-mediated strand displacement (TMSD) is a process in which one of the strands within a nucleic acid duplex is exchanged for another: displacement of the incumbent strand is initiated by the binding of the invader to a short single-stranded toehold region on the complementary strand \cite{Zhang_2009,Yurke2000} (Fig.\ \ref{fig:tmsd}(a)). TMSD has many applications in nanotechnology, including in the construction of synthetic molecular circuits \cite{Qian2011}. DNA-RNA hybrid TMSD is of particular interest by virtue of its relevance to \emph{in vivo} applications \cite{Liu_2021}. Strand displacement has also been argued to play important roles in various naturally-occurring RNA systems \cite{Hong2019}.

We note that oxDNA has been remarkably successful in reproducing experimental observations related to TMSD, having been used, for example,  to study mismatches as a tool for modulating strand displacement kinetics\cite{Machinek2014,Haley2020}. RNA strand displacement has likewise been simulated using the oxRNA model \cite{_ulc_2015}.

Here, we use our newly developed model to study strand displacement systems involving DNA-RNA hybrids. As in the melting simulations, we use a combination of VMMC and umbrella sampling to explore the state space efficiently. From the simulations, we obtain unbiased estimates of equilibrium populations of states parameterised by the number of substrate-invader hydrogen bonds. In simulations of toehold-mediated strand displacement, we assigned a weight of zero to states with no hydrogen bonds between substrate and invader or substrate and incumbent hydrogen bonds, to prevent dissociation. Umbrella sampling weights were chosen (by trial and error) so that all states have approximately equal occupancy (within an order of magnitude) in the biased ensemble. Assuming that the state space has been adequately sampled, the free energy difference $\Delta G$ between states $A$ and $B$ can be written as

\begin{equation}
    G(A) - G(B) = -k_B T \ln \left[\frac{p(A)}{p(B)}\right],
\end{equation}
where $p(A)$ and $p(B)$ are the probabilities of being in state $A$ and $B$ respectively. We can similarly compute free energy profiles for systems with multiple states. For every system studied, we ran 10 independent simulations for $10^9$ time-steps each, at 37\,°C and a 0.5\,M monovalent salt concentration using the average-sequence model. We simulated four systems---an RNA strand invading dsDNA, a DNA strand invading dsRNA, a DNA strand invading a hybrid duplex to displace an RNA incumbent from a DNA substrate, and finally an RNA strand invading a hybrid duplex to displace an DNA incumbent. In each case, the toehold region was 4 nucleotides long, with a 10-nucleotide branch migration domain. Results are shown in Fig.\ \ref{fig:tmsd}(b).

A common feature of all of the free energy profiles is the initial downhill trajectory in the range of 1 to 4 invader-substrate hydrogen bonds. This is associated with toehold binding, which is always favourable, as there is no competition between strands. Generally, there is an entropic barrier associated with the formation of a branch junction during strand displacement, which is seen as an activation barrier in the branch migration region (for RNA invading dsDNA and DNA invading hybrid). In the case of DNA invading dsRNA, the landscape is steeply uphill, as on average dsRNA is substantially more thermally stable than a DNA-RNA hybrid. Conversely, when RNA invades a hybrid this results in the formation of dsRNA, which is much more thermally stable than a hybrid duplex, resulting in a downhill landscape.  This can be understood in terms of the difference in average melting temperature between dsDNA and dsRNA---around 60\,°C and 71\,°C for a 10 base-pair duplex respectively The difference between the free energy landscapes for RNA invading dsDNA and DNA invading a hybrid is more subtle because hybrids and dsDNA are quite close in melting temperature (around 61\,°C for a 10 base-pair hybrid duplex).  It is likely that this relative difference is smaller than the typical effects of varying base sequence.

The simulations performed here only scratch the surface of what can be studied with the model---future work will investigate the effect of sequence on TMSD free energies and kinetics. Preliminary simulations with the model suggest that free energy landscapes, as well as reaction kinetics, are strongly sequence-dependent. We are also looking into how secondary structure in the RNA strand impacts the reaction. Given the success of previous oxDNA models in studying TMSD, we are confident that our DNA-RNA hybrid model will provide useful insights.

\subsection{\label{sec:level2}R-loop resolution}


\noindent An R-loop is a three-stranded nucleic acid structure consisting of double-stranded DNA which is partially hybridised to complementary RNA. As discussed in Section I, this is possibly the most important naturally occurring DNA-RNA hybrid system. 

\begin{figure}[h!]
\includegraphics[scale=0.082]{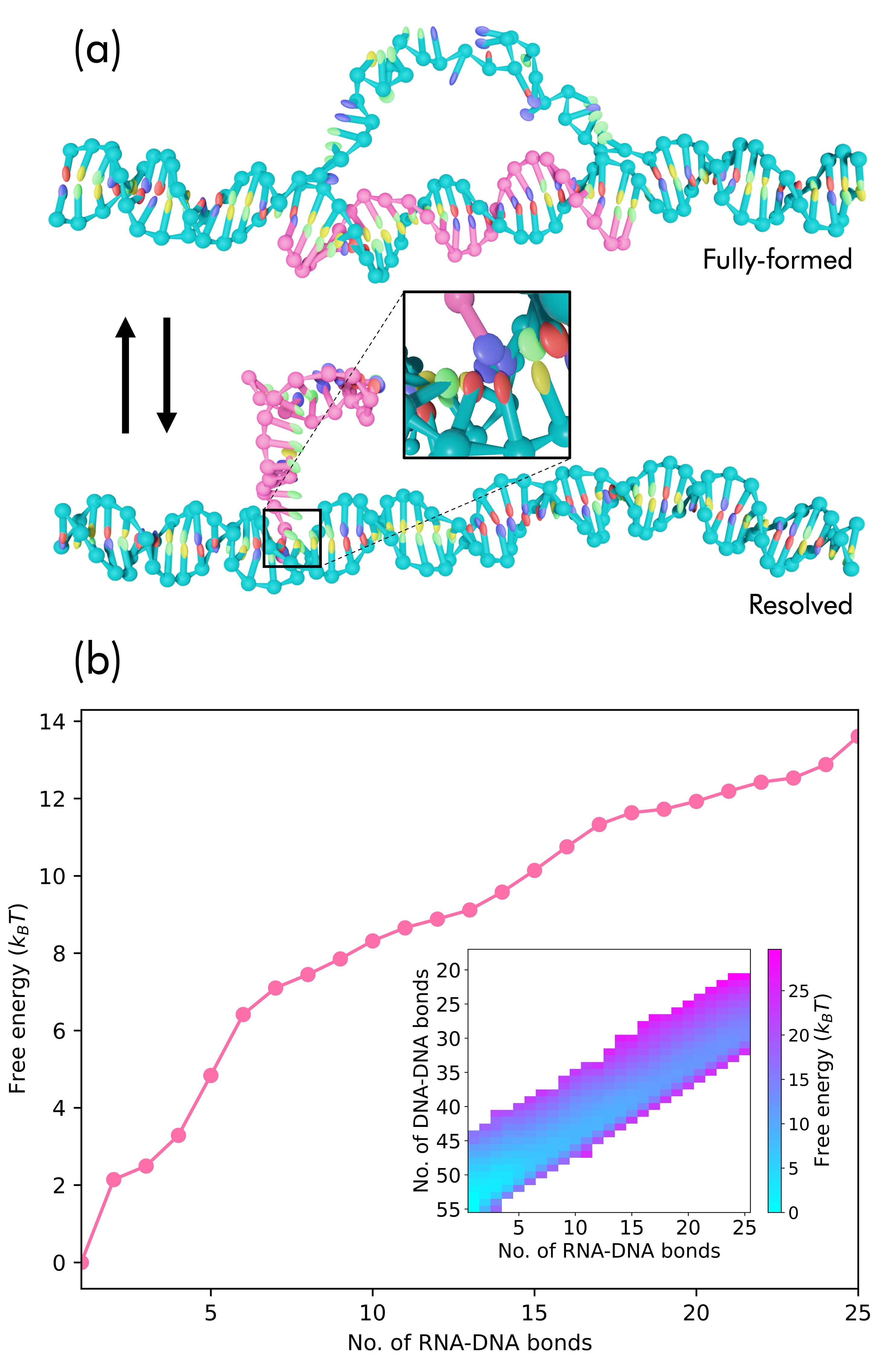}
\caption{\label{fig:rloop} Studying the resolution of a short R-loop. (a) A fully-formed 25-nucleotide R-loop, consisting of double-stranded DNA (blue) and a single strand of RNA (pink). Through the process of strand displacement, the system can resolve the R-loop by forcing out the RNA strand. In our simulations, this transition is sampled many times in both directions. (b) Free energy of the system as a function of the number of RNA-DNA hydrogen bonds and (inset) the number of both DNA-DNA and RNA-DNA bonds (states with fewer than a total of approximately 45 base pairs are not sampled).}
\end{figure}

We use our coarse-grained model to simulate the resolution of an R-loop. While this system appears to be similar to the TMSD studied in Section IV A, as both involve DNA-RNA strand displacement, we observe behaviour which is quite different. The simulation protocol used closely resembles our TMSD simulations. We study a single R-loop consisting of 55 base-pair double-stranded DNA which is hybridised to a 25-nucleotide RNA strand at its centre (Fig.\ \ref{fig:rloop} (a), top). As before, in order to prevent strand dissociation we restrict the system to states with at least one DNA-DNA and one RNA-DNA hydrogen bond, and use average-sequence parameters. In this case, we ran separate simulations for two overlapping windows of the order parameter space---one restricted to 1--13 RNA-DNA hydrogen bonds, and another to 13--25 bonds. We performed 10 independent VMMC simulations per window, each for $3\times10^8$ time-steps. Temperature and monovalent salt concentration were the same as for our TMSD simulations.

Computed free energy profiles are shown in Fig.\ \ref{fig:rloop}(b). There is a barrier of around $2\,k_B T$ associated with the transition from 1 to 2 RNA-DNA bonds. The zoomed-in snapshot of the resolved state in Fig.\ \ref{fig:rloop}(a) suggests an explanation. In the resolved state, the DNA double helix tends to be fully closed, with the RNA strand forming a weak hydrogen bond with one of the DNA strands. As a result, in order to make the transition from one to two RNA-DNA bonds, two DNA-DNA bonds must be broken, which is energetically costly. This is in part an artefact of restricting the simulation to bound states. Without this restriction, the RNA strand would have dissociated completely in the resolved state. 

In general, we observe that the formation of the DNA-RNA hybrid in this particular system is significantly less favourable than in the analogous TMSD reaction of RNA invading dsDNA, depicted in Fig.\ \ref{fig:tmsd}(b). Several factors contribute to the difference between the two energy landscapes. In a fully-formed R-loop, displacement of the RNA strand can take place from either end: the DNA loop is tethered at both sides, increasing its proximity to the hybrid, making displacement more likely. Resolving an R-loop is clearly entropically favourable, as it entails exchange of a single strand tethered at both ends for one tethered at only one end in our simulations, or fully displaced in practice, thus having much greater conformational freedom. 

We also observe an oscillatory component to the free energy which has minima at R-loop sizes of around 13 and 23 RNA-DNA bonds. When the DNA-RNA hybrid helix is of a size roughly commensurate with its pitch (around 11 base pairs), the ends of the displaced DNA loop are on the same side of the duplex, which entails higher conformational freedom. Conversely, at half a turn away, e.g.\ around 18, the ends are at opposite sides of the duplex, reducing conformational freedom, and leading to a slight additional increase in free energy cost.

The inset in Fig.\ \ref{fig:rloop}(b) depicts a 2D free-energy landscape that provides additional information about the system. The presence of the R-loop destabilises the DNA double helix beyond the region of the DNA-RNA hybrid, with states which are not fully hybridised being readily accessible. This is clear from the fact that, at any given number of RNA-DNA bonds, states with numbers of DNA-DNA bonds below what would be expected for a fully hybridised system (55 bonds in total) are sampled.

The stability of an R-loop depends on its length and sequence\cite{Landgraf1995}. An obvious future application of our model would be a comprehensive study of the effects of these factors. The kinetics of R-loop resolution could also be studied using specialised sampling techniques. 

\begin{figure*}
\includegraphics[scale=0.13]{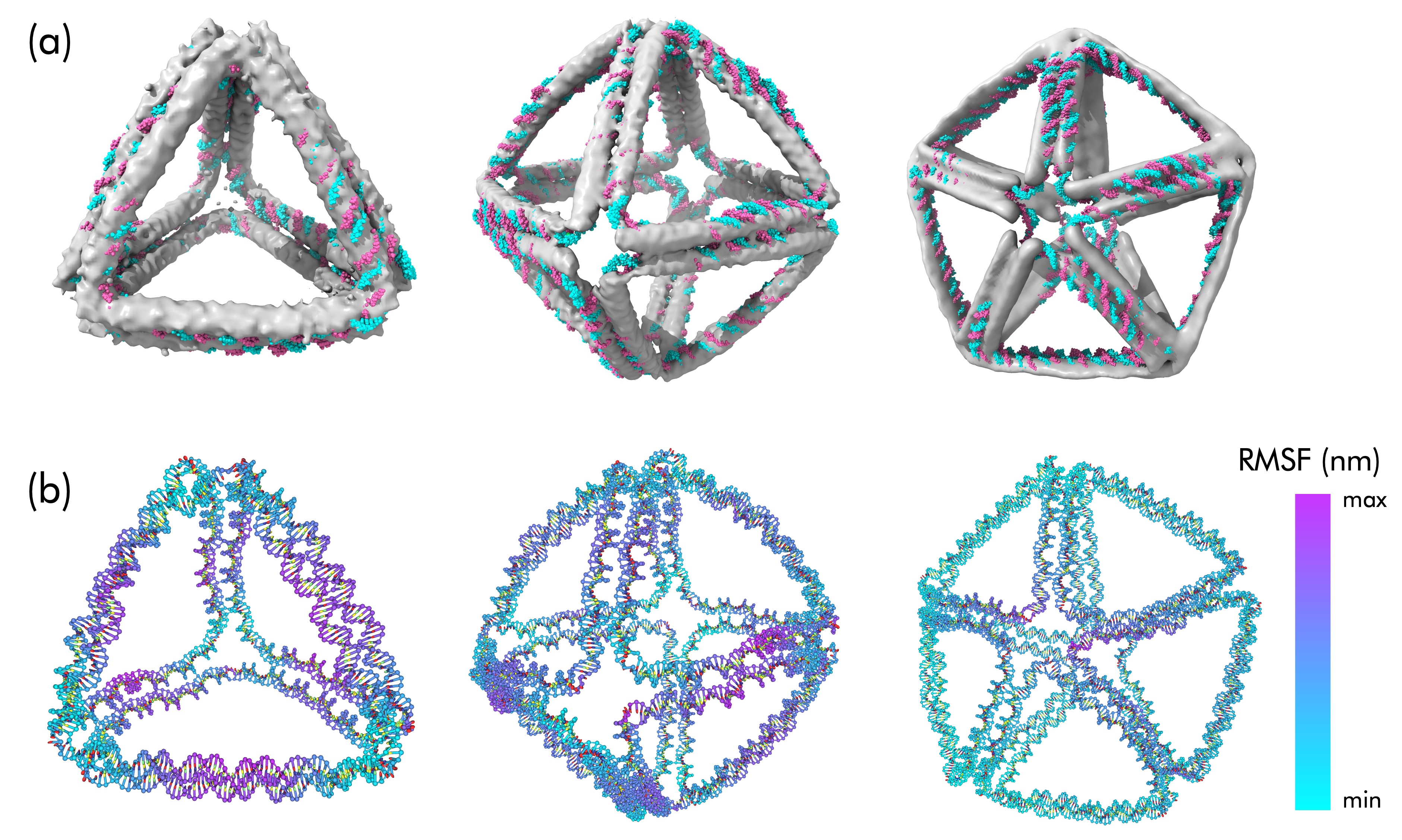}
\caption{\label{fig:origami} Mean structures of RNA-scaffolded origami simulated using the model. (a) Atomic models of the tetrahedron (left), octahedron (middle) and pentagonal bipyramid (right), each consisting of an RNA scaffold strand (pink) and DNA staples (blue). Experimentally obtained cryo-EM densities (grey) have been superimposed onto each structure. (b) Structures with colouring to indicate the per-nucleotide RMSF. The structures have different fluctuation ranges: 1.15--1.72\,nm, 1.38--2.13\,nm and 1.47--3.87\,nm respectively.}
\end{figure*}
\subsection{\label{sec:level2}RNA-scaffolded wireframe origami}

\noindent Nucleic acid origami is one of the most common techniques used for assembling single-stranded DNA/RNA building blocks into a target structure. Origami nanostructures consist of a scaffold, which is a long strand running through the entire assembly, and shorter staple strands which hybridise to two or more scaffold domains to control its spatial arrangement. Domains of the scaffold strand which are widely separated in its primary sequence can be held in close spatial proximity in the final structure. This technique has been applied primarily to DNA, although interest in the design of DNA-RNA hybrid nanostructures is increasing.

We have used our model to simulate three hybrid wireframe origami nanostructures from Parsons \textit{et al}.\ \cite{Parsons_2023} which consist of an RNA scaffold and DNA staples. The structures were designed assuming a double helix with a pitch of 11 base-pairs per turn, which is roughly reproduced by our model. We performed MD simulations at 4\,°C and a monovalent salt concentration of 0.3\,M, to match the experiments.  Each structure was simulated for $10^7$ time-steps and the positions of particles were sampled every $10^4$ time-steps for analysis. We simulated three nanostructures---a tetrahedron, an octahedron and a pentagonal bipyramid---each having edges 66 base-pairs long. For each, we calculated the mean structure and per-nucleotide RMSF (root-mean-square fluctuation). From these mean structures, we reconstructed all-atom models of the nanostructures using the oxDNA-to-PDB converter on TacoxDNA~\cite{Suma_2023} (by superimposing atomic coordinates onto individual nucleotides) and then aligned them with cryo-EM densities, obtained by Parsons \textit{et al}.\ and retrieved from EMDB\cite{Lawson2015}, using ChimeraX~\cite{Pettersen2020}.

Fig.\ \ref{fig:origami} compares our results to the experimental data. Our model captures the measured structures reasonably well, with no systematic strain build-up. For the tetrahedron and octahedron it is immediately clear that structural fluctuations are concentrated at edge centres.

\section{\label{sec:level1}Conclusion} 
\noindent We have introduced a new coarse-grained model, based on existing oxDNA and oxRNA models, which enables the simulation of DNA-RNA hybrids.  As with previous models, we parameterised the hydrogen bonding interaction to reproduce the melting temperatures of short duplexes. Quantitative agreement with the experimentally-calibrated nearest-neighbour model of the thermodynamics of hybrid duplexes is nearly as close as that achieved for DNA and RNA duplexes using  oxDNA and oxRNA. The persistence length and stretching modulus derived from simulations of longer duplexes are consistent with experimental values, although some uncertainty about their values remains.   The conformation of DNA-RNA hybrid duplexes is a compromise between the structures preferred by DNA and RNA alone. As a result, stabilization of the duplex by stacking interactions is reduced, necessitating the increase of hydrogen bonding strength to produce desired melting temperatures. One consequence of this choice is that the model overestimates the stability of short double-stranded helices---something which users of the model should keep in mind.  Nevertheless, the overall performance of our DNA-RNA hybrid model for the systems we studied gives us confidence that it will be able to capture sequence-dependent kinetics/thermodynamics of more complex biophysical processes.  A future version of the model will include a modified stacking potential which can accommodate the preferred conformations of dsDNA, dsRNA and DNA-RNA hybrids.

We have demonstrated the versatility and applicability of our model by performing simulations for three different systems.  Our study of toehold-mediated strand displacement using the average-sequence model suggests that the relative stabilities of DNA-DNA, RNA-RNA and DNA-RNA duplexes plays a key role in determining the free energy landscapes of hybrid displacement reactions.  Our simulations show that the biophysics of R-loop resolution includes geometric effects related to the commensurability of the R-loop length and the pitch of the double helix. Finally, we have shown that our model can help validate DNA-RNA hybrid origami designs.

Future work will focus on DNA-RNA hybrid systems at time and length-scales that are  inaccessible to all-atom simulations, including the sequence-dependent kinetics of strand displacement reactions and the effects of RNA secondary structure motifs. 

\section{\label{sec:level1}Code Availability} 
\noindent The code implementing the model alongside the supporting documentation can be found at \url{https://lorenzo-rovigatti.github.io/oxDNA/}. A new topology file format supporting DNA-RNA hybrids has been implemented in the official oxDNA code, and the accompanying suite of analysis tools has likewise been extended to enable the analysis of systems containing both DNA and RNA. The online visualization tool oxView.org \cite{Bohlin2022} has been extended to also support viewing of DNA-RNA hybrids. The simulations performed here were run on single CPUs, although a GPU version of the model is a likely future development.  

\section{\label{sec:level1}Acknowledgments} 
\noindent The authors thank Thomas Ouldridge and Jonathan Bath for useful discussions, Lorenzo Rovigatti and Erik Poppleton for their help with code development, and Erik Winfree for suggesting the name oxNA for the model. E.J.R. acknowledges financial support provided by the Clarendon Fund, Somerville College (Oxford) and the Engineering and Physical Sciences Research Council (grant No.\ EP/W524311/1). We also thank the Advanced Research Computing service, University of Oxford, for computer time. P.\v{S}.~acknowledges support by the National Science Foundation under grant No.\ CCF 2211794.

\section{\label{sec:level1}References} 
\bibliography{aipsamp}

\end{document}